# Distinct citation distributions complicate research evaluations. A single indicator that universally reveals research efficiency cannot be formulated

Alonso Rodríguez-Navarro

*Departamento de Biotecnología-Biología Vegetal, Universidad Politécnica de Madrid, Avenida Puerta de Hierro 2, 28040, Madrid, Spain*

E-mail address: alonso.rodriguez@upm.es. Telephone number: +34 916333869

ORCID 0000-0003-0462-223X

**Abstract**

**Purpose**: Analyze the diversity of citation distributions to publications in different research topics to investigate the accuracy of size-independent, rank-based indicators. Top percentile-based indicators are the most common indicators of this type, and the evaluations of Japan are the most evident misjudgments.

**Design/methodology/approach**: The distributions of citations to publications from countries and in journals in several research topics were analyzed along with the corresponding global publications using histograms with logarithmic binning, double rank plots, and normal probability plots of log-transformed numbers of citations.

**Findings**: Size-independent, top percentile-based indicators are accurate when the global ranks of local publications fit a power law, but deviations in the least cited papers are frequent in countries and occur in all journals with high impact factors. In these cases, a single indicator is misleading. Comparisons of proportions of uncited papers are the best way to predict these deviations.

**Research limitations**: The study is fundamentally analytical; its results describe mathematical facts that are self-evident.

**Practical implications**: Respectable institutions, such as the OECD, European Commission, US National Science Board, and others, produce research country rankings and individual evaluations using size-independent percentile indicators that are misleading in many countries. These misleading evaluations should be discontinued because they cause confusion among research policymakers and lead to incorrect research policies.

**Originality/value**: Studies linking the lower tail of citation distribution, including uncited papers, to percentile research indicators have not been performed previously. The present results demonstrate that studies of this type are necessary to find reliable procedures for research assessments.

# 1. Introduction

Research evaluation of countries and institutions is one of the most important applications of bibliometrics (Garfield & Welljams-Dorof, 1992). Without research evaluations, research policy is arbitrary. Although the product of research systems—the progress of knowledge—is intangible, in other aspects, research systems are similar to those producing typical merchandise. In research systems, the absence of research evaluations would be equivalent to having an industry in which the production department is disconnected from the sales department. In this anomalous case, the production department would be developing new products, ignoring whether they are sold or stacked in a warehouse, and then disposing of them because they are not sold.

Research systems can be assessed in terms of either size or quality. Regarding size, the method may be as simple as counting the number of publications, but regarding quality, the assessment is a much more difficult task. At a low level of aggregation, research assessments of quality should be performed by peer review, but at the country level, the dimensions make it impossible to organize peer review assessments. Even in institutions, a comprehensive peer review may be very difficult (Martin, 2011). In these cases, bibliometric assessments are the best solution (Abramo et al., 2013; Abramo et al., 2019).

Bibliometric evaluations have been performed for a long time (Godin, 2006; Leydesdorff, 2005), boosted by a pioneering study of Francis Narin (Narin, 1976). Regretfully, in too many cases, research success has been defined 'operationally' as simply amounting to the score of the proposed index (Harnad, 2009). With significant progress in the current century (Aksnes et al., 2019; Tahamtan & Bornmann, 2019; Waltman, 2016), many studies have demonstrated that citation-based metrics are the most convenient indicators for research evaluations (Aksnes et al., 2019; Waltman, 2016) if they are used at a high aggregation level (Aksnes et al., 2023; Thelwall et al., 2023).

A great number of citation-based parametric and nonparametric indicators have been proposed in the last 20–30 years (van Noorden, 2010; Wildgaard et al., 2014). Among them, top percentile indicators (Bornmann et al., 2013; Waltman & Schreiber, 2013) are preferred by the most well-known international or national institutions. Some examples are the OECD, US National Board, European Commission, National Institute of Science and Technology Policy in Tokyo, and the CWTS of the University of Leiden. These indicators are accurate and reliable if the results at different percentiles fit a power law (Rodríguez-Navarro & Brito, 2019). In other cases, they fail. Japan is a good example of these failures because it is a country with a high scientific level that top percentile and other indicators suggest is a research-developing country (Pendlebury, 2020).

The power law that percentile and other rank-based indicators follow represents an ideal model. Similar to how real gases frequently deviate from the ideal gas law, the research outputs of some institutions and countries depart from the ideal model. These deviations give rise to uncertain rankings of countries and institutions because Japan need not be an isolated case of misjudgment (Rodríguez-Navarro, 2024b).

In addition to these deviations, or perhaps as the basis of these deviations, research may be addressed to produce either incremental innovations or scientific advancements, which have different citation practices (Rodríguez-Navarro & Brito, 2022). Consequently, studies focused on understanding the bibliometric differences between these two types of research are at the forefront of scientometric research.

*1.1. Wrong diagnoses and misguided research policies*

The notion that a suitable research policy must be based on reliable research assessments is rational and empirical. A well-known example of a wrong diagnosis of research success that misguides research policy is the EU. This failure was described 18 years ago (Dosi et al., 2006), and subsequent publications have demonstrated its permanence. First, the EU's research policy was dominated by the "European paradox" (Albarrán et al., 2010), and later, when many academic publications criticized the paradox, the same idea was expressed by the catchphrase: "Europe is a global scientific powerhouse" (Rodríguez-Navarro & Brito, 2020b). In contrast with this political propaganda of success, the analysis of the most cited papers shows that, in terms of contribution to pushing the boundaries of knowledge, the EU is well behind the USA and China in most technological fields (Rodríguez-Navarro, 2024a).

The case of the EU is only one example; likely, the same problem affects many other countries. The comparison between Norway and Singapore is another example. These two countries have a similar number of inhabitants, and in both countries, the GDP per capita is very high, which implies that research is not restricted by economic difficulties. Despite these similarities, the analysis of the global ranks of the most cited papers in technological topics shows the overwhelming superiority of Singapore (Rodríguez-Navarro, 2024a). This implies that the research success suggested by some bibliometric analyses (Sivertsen, 2018) does not exist and that, in comparison to Singapore, Norwegian research policy is far from being successful.

A wrong diagnosis can occur in both senses, overstating or understating the real success. In contrast with the overestimations described above, Japan is trying to boost the global ranking of its universities (Normile, 2024), but its research level is probably much better (Rodríguez-Navarro, 2024a) than suggested by common indicators (Pendlebury, 2020), even when they are obtained in Japan (National Institute of Science and Technology, 2022).

To sum up, accurate research assessments are necessary for accurate research policies and successful contributions to scientific and technological progress.

*1.2. Global ranks and deviations from the ideal power law*

In the ideal model described above (Section 1), when global and local publications are ordered from highest to lowest number of citations, local versus global ranks fit a power law (Rodríguez-Navarro & Brito, 2018). Plotting this relationship on double-logarithmic scales produces a straight line where deviations are easily detected. In ideal cases there is an easily testable property because $P_{top\ 10\%}/P$, $P_{top\ 5\%}/P_{top\ 50\%}$, and $P_{top\ 1\%}/P_{top\ 10\%}$ are equal (henceforth, I will use the Leiden Ranking notation; $P_{top\ x\%}$ means the number of papers in the top x% by citation), and the number of papers in all top percentiles can be easily calculated from P and the number in a single top percentile (Rodríguez-Navarro & Brito, 2021). The issue is that the publications of many research systems do not fit the ideal model (Rodríguez-Navarro, 2024b).

Country rankings published by the OECD (2016), European Commission (2022), Japan (National Institute of Science and Technology, 2022), France (Hcéres, 2019), and other international and national institutions use the share of the top 10% most-cited scientific publications as a measure of scientific excellence. This would seem like a reasonable method because the top percentiles have been validated by correlation against the results of the UK Research Excellence Framework, which are based on peer review (Rodríguez-Navarro & Brito, 2020a; Traag & Waltman, 2019). However, it would be hard to find a senior researcher who believes that one in 10 scientific publications makes a significant contribution to the advancement of science. Therefore, a high share of the top 10% most-cited scientific publications could be taken as success in normal research but not in revolutionary research (Kuhn, 1970). The revolutionary research that pushes the boundaries of knowledge must be evaluated in much narrower percentiles, in the range of the 0.01-0.02% most-cited papers (Bornmann et al., 2018; Poege et al., 2019). However, as just mentioned above, if research outputs fitted the ideal model, evaluations based on the top 10% most-cited papers could be used to evaluate breakthroughs that are 1,000 times less frequent (Rodríguez-Navarro & Brito, 2019), but only in these cases.

*1.3. Scientometric challenges*

Research has diverse impacts on society (Bornmann, 2013; Greenhalgh et al., 2016; Martin, 1996), which leads to the conclusion that research cannot be assessed using a single indicator. In contrast, the effect of research on science could be described exclusively as its contribution to the progress of knowledge. In principle, with this restriction, a single indicator of research might be sufficient to describe the efficiency of a research system. However, this is not entirely correct because research can be aimed at two different objectives: boosting incremental innovations or pushing the boundaries of knowledge (Section 1). These two types of research have different citation practices, with citation being less frequent in technical research, and are mixed in different proportions in most countries and institutions. Furthermore, citation-based rank analyses (e.g., top percentile indicators) are based on the comparison of global and local citation distributions, and a reasonable conjecture is that the ideal rank power law appears when the proportions of these two types of research in the country or institution under evaluation is similar to the proportions in global research. Japan does not fulfill this requirement, and for this reason, its evaluations fail (Rodríguez-Navarro, 2024b). The same can occur in other countries and institutions.

To overcome these difficulties, the evaluation can be focused on the contribution to pushing the boundaries of knowledge at citation levels that correspond only to this type of research. Then, the evaluations can be performed with a single indicator (Rodríguez-Navarro & Brito, 2024). However, even in this case, the challenge is not solved. An indicator based on the most cited papers is size-dependent and provides no information about the efficiency of the system, which implies that the comparison of countries or institutions of different sizes is impossible. In other words, the indicator does not convey to what extent successful achievements stem from size or efficiency.

In summary, it is not clear how to obtain comprehensive research evaluations that provide solid information to research policymakers. This drawback does not exist when the ideal rank power law applies, but this does not occur universally.

*1.4. Aim of this study*

The uncertainty regarding the accuracy of research indicators directly calculated from easily measured statistical data, such as $P_{top\ 10\%}/P$ or $P_{top\ 1\%}/P$, prompted a study of the relationships between citation distributions and the accuracy of these indicators. For this purpose, this study assumes two ideal models: a lognormal distribution for citations and a power law relationship between global and local ranks of papers, and investigates the deviations from these ideal models. Deviations in the upper tail are important regarding the contribution to pushing the boundaries of knowledge, but they have been studied (Rodríguez-Navarro, 2024b; Rodríguez-Navarro & Brito, 2024). Currently, the most important challenge is in the lower tail and the part of the citation distribution with papers that are not highly cited.

This study uses research topics, which implies more homogeneous populations of papers than fields, which aggregate many topics. It has two parts: the first investigates the deviations from the ideal model at global and country levels, while the second is focused on journals in order to study more homogeneous and diverse populations of papers than in the cases of countries or institutions. It was conjectured that journals' publications should facilitate the analysis of indicators and the study of extreme cases. It is unlikely that the publications of a country or institution have a global success equivalent to that of journals such as *Nature* or *Science*, but a comprehensive description should also consider these extreme cases.

## 2. Methods

This study is based on citation data obtained from Clarivate Web of Science, as described in a previous paper (Rodríguez-Navarro, 2024b), using the same publication (2014–2017) and citation (2019–2022) windows, as well as domestic publications. The recorded journal impact factors (JIF) correspond to 2019. In the searches, the topic referred to in the text as "solar cells" also includes "photovoltaics," and the case referred to as "dementia" also includes "Parkinson" and "Alzheimer."

To construct the rank plots, either for countries or journals, the global papers were ordered by the number of citations from highest to lowest, with 1 assigned to the most cited paper. Because many papers have the same number of citations, to construct more accurate rankings, in addition to the number of citations, the papers were subsequently ordered by the publication year, the average number of citations per year, and the DOI. To obtain the global ranks of country papers, these papers were ordered as for global papers, and the papers in this list were identified in the global list together with their global ranks. To find the global ranks of journals' papers, the global list was ordered alphabetically by the name of journals and segmented for each journal.

The logarithmic bins of the citation distributions are shown in the histograms, which also include the number of papers with zero, one, and two citations. Normal probability plots of the log-transformed number of citations (or citations plus 1, if in the list there are papers with zero citations) were constructed as described by Boylan and Cho (Boylan & Cho, 2012). The plotting positions were obtained using the formula $p_i = (i - 0.5)/n$, where $n$ is the number of papers. The results of the Kolmogorov-Smirnov goodness-of-fit tests of the log-transformed numbers of citations were obtained at https://contchart.com/goodness-of-fit.aspx.

The proportion of lowly cited and uncited papers in the citation distribution of countries and journals has special interest in this study. These papers may belong to a specific type that is abundant in highly technological countries (Rodríguez-Navarro &

Brito, 2022), but this relationship has not been specifically investigated. The proportion of uncited papers is probably the best indicator for the whole population of papers of this type, and in this sense of indicator of a population of papers is used throughout this study.

## 3. Results

### 3.1. Deviations from the rank power law: countries

The proportion of uncited papers is notably variable among research topics (Rodríguez-Navarro, 2024b). In the 10 technical and biomedical topics studied here, the proportion of uncited papers varies from 11% for semiconductors to 3% in lithium batteries and dementia (Table 1). Despite this variability, the normal probability plots of the log-transformed number of citations (plus 1) reveal very similar inflated lower tails compared to the rest of the papers. Figure 1 depicts the plots for lithium batteries and semiconductors.

Table 1. Number of papers and proportion of uncited papers in selected research topics.

| Topic | Number | MNC[a] | Uncited (%) |
|---|---|---|---|
| Semiconductors | 58393 | 18.4 | 10.8 |
| Steel | 69128 | 13.8 | 8.9 |
| Concrete | 34126 | 17.1 | 7.9 |
| Solar cells | 61202 | 22.3 | 7.6 |
| Combustion | 38403 | 16.7 | 5.9 |
| Immunity | 42586 | 21.8 | 4.4 |
| Stem cells | 86647 | 20.9 | 4.3 |
| Graphene | 82757 | 29.9 | 3.8 |
| Dementia | 39767 | 22.6 | 3.2 |
| Lithium batteries | 32318 | 32.4 | 3.1 |

[a] Mean Number of Citations

Confirming a previous study (Rodríguez-Navarro, 2024b), it is found that a notable number of countries deviate from the ideal model ($P_{top\ 10\%}/P = P_{top\ 5\%}/P_{top\ 50\%} = P_{top\ 3\%}/P_{top\ 30\%} = P_{top\ 1\%}/P_{top\ 10\%}$), which indicates that research country rankings based on a single indicator, either P$_{top\ 10\%}$/P or P$_{top\ 1\%}$/P, can be misleading depending on the selected countries. Tables 2 and 3 show the results for 16 countries in the research topics of graphene and solar cells. Deviations of P$_{top\ 1\%}$/P$_{top\ 10\%}$ exclusively indicates a deviation in the extreme of the upper tail that might affect to a very low proportion of papers (Rodríguez-Navarro, 2024b). Excluding this ratio, it can be concluded that, roughly, the number of countries that significantly deviate from the ideal model is around 50% or even more.

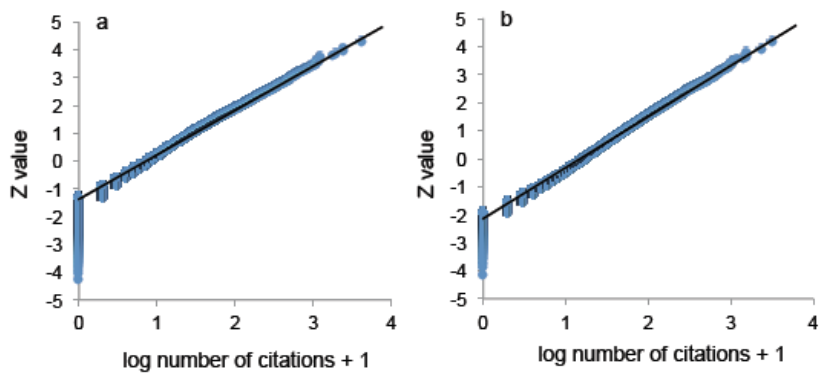

Figure 1. Normal probability plots of the log-transformed number of citations plus 1 of publications in the topics of semiconductors (a) and lithium batteries (b)

Table 2. Basic description of publications on graphene from selected countries, including percentile ratios.

| Country | P[a] | P$_0$[b] (%) | MNC[c] | P$_{top\ 10\%}$/P | P$_{top\ 5\%}$/P$_{top\ 50\%}$ | P$_{top\ 3\%}$/P$_{top\ 30\%}$ | P$_{top\ 1\%}$/P$_{top\ 10\%}$ |
|---|---|---|---|---|---|---|---|
| China | 35493 | 3.6 | 31.8 | 0.104 | 0.102 | 0.098 | 0.088 |
| EU | 7555 | 7.1 | 22.3 | 0.059 | 0.080 | 0.088 | 0.107 |
| USA | 5784 | 5.1 | 42.1 | 0.127 | 0.166 | 0.194 | 0.221 |
| South Korea | 4701 | 5.2 | 24.1 | 0.078 | 0.069 | 0.065 | 0.054 |
| India | 3794 | 3.1 | 21.1 | 0.049 | 0.034 | 0.023 | |
| Japan | 1788 | 10.3 | 17.0 | 0.041 | 0.094 | 0.108 | 0.095 |
| Germany | 843 | 4.2 | 21.4 | 0.064 | 0.097 | 0.102 | |
| Singapore | 683 | 2.3 | 47.8 | 0.182 | 0.175 | 0.163 | 0.121 |
| UK | 666 | 2.7 | 27.1 | 0.083 | 0.078 | 0.095 | |
| Italy | 664 | 3.6 | 18.1 | 0.039 | 0.039 | | |
| Spain | 657 | 2.3 | 22.4 | 0.055 | 0.055 | 0.053 | |
| Australia | 552 | 2.0 | 47.9 | 0.199 | 0.209 | 0.213 | 0.155 |
| Canada | 514 | 3.7 | 29.5 | 0.097 | 0.088 | 0.092 | |
| France | 410 | 5.9 | 21.4 | 0.066 | 0.067 | | |
| Switzerland | 136 | 6.6 | 26.4 | 0.088 | 0.127 | | |
| Netherlands | 132 | 2.3 | 34.0 | 0.136 | 0.131 | | |

[a] Number of publications; [b] uncited papers; [c] Mean Number of Citations

Table 3. Basic description of publications on solar cells from selected countries, including percentile ratios.

| Country | P[a] | P$_0$[b] (%) | MNC[c] | P$_{top\ 10\%}$/ P | P$_{top\ 5\%}$/ P$_{top\ 50\%}$ | P$_{top\ 3\%}$/ P$_{top\ 30\%}$ | P$_{top\ 1\%}$/ P$_{top\ 10\%}$ |
|---|---|---|---|---|---|---|---|
| China | 12806 | 7.7 | 18.8 | 0.086 | 0.085 | 0.083 | 0.064 |
| EU | 8170 | 5.8 | 18.4 | 0.082 | 0.066 | 0.056 | 0.048 |
| USA | 5471 | 4.6 | 32.6 | 0.146 | 0.141 | 0.139 | 0.141 |
| South Korea | 4188 | 13.9 | 17.8 | 0.066 | 0.077 | 0.087 | 0.104 |
| India | 3118 | 5.8 | 14.1 | 0.051 | 0.040 | 0.035 | |
| Japan | 2727 | 12.3 | 13.8 | 0.059 | 0.079 | 0.095 | 0.093 |
| Germany | 1829 | 6.4 | 17.4 | 0.078 | 0.080 | 0.076 | 0.063 |
| Italy | 1161 | 4.0 | 20.5 | 0.091 | 0.053 | 0.034 | |
| UK | 890 | 4.3 | 43.1 | 0.179 | 0.176 | 0.204 | 0.208 |
| Australia | 803 | 4.7 | 24.0 | 0.130 | 0.102 | 0.081 | |
| Spain | 794 | 5.3 | 19.1 | 0.081 | 0.060 | 0.051 | |
| France | 673 | 5.9 | 15.0 | 0.062 | 0.044 | 0.048 | |
| Canada | 546 | 6.0 | 24.8 | 0.106 | 0.084 | 0.083 | |
| Singapore | 363 | 4.7 | 27.1 | 0.132 | 0.129 | 0.140 | 0.167 |
| Switzerland | 345 | 3.5 | 49.3 | 0.209 | 0.251 | 0.248 | 0.167 |
| Netherlands | 304 | 3.3 | 21.1 | 0.099 | 0.083 | 0.063 | |

[a] Number of publications; [b] uncited papers; [c] Mean Number of Citations

When deviations from the ideal model are small, the double rank plots show that countries' comparisons are accurate. For example, this is the case in the comparisons of Spain and Singapore in graphene and the USA and Germany in solar cells (Figure 2). In these cases, differences in size should not be an impediment to finding an accurate indicator of efficiency that, along with the difference in size, clearly defines the research differences between the two countries. In contrast, when the deviations from the ideal system are large, finding indicators that reveal the differences in size and efficiency of two countries seems very difficult. This is the case in the comparison of Japan and India in graphene (Table 2; Figure 3). Depending on the publications considered, the comparative judgment of research efficiencies (from the slope of the fitted straight lines) changes depending on the rank range considered. Considering the 50% least cited papers, India seems to be more efficient than Japan, but using the 10% most cited papers, the conclusion is the opposite. More importantly, if we calculate the efficiency from the 10% most cited papers, we would assess the size of the research system in India as bigger than it is, while in Japan the assessment would be smaller than it is.

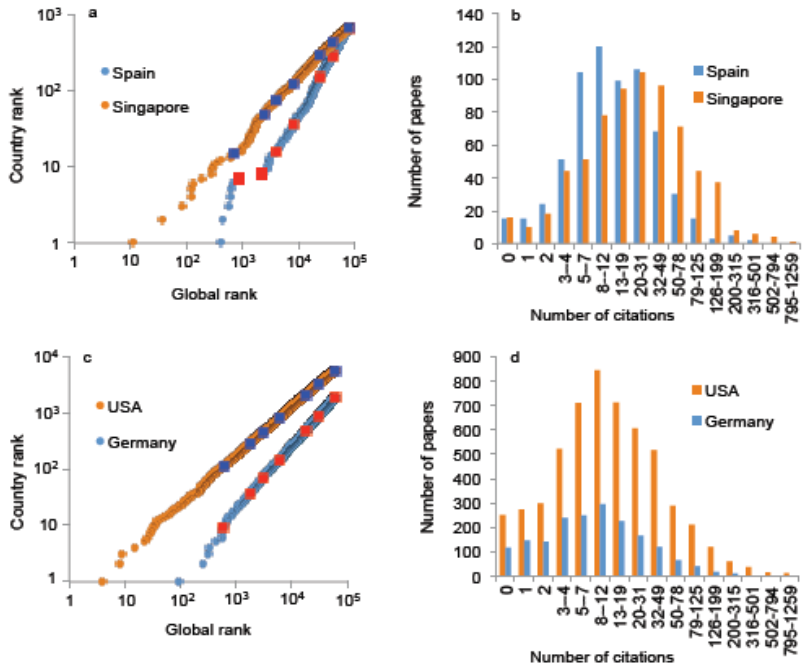

Figure 2. Double rank plots (a, c) and citation distributions (b, d) of the publications from Spain and Singapore in graphene (a, b) and the USA and Germany in solar cells (c, d). Square symbols in (a) and (c) represent the position of papers in top percentiles: 1, 3, 5, 10, 30, 50, and 100, from bottom to top.

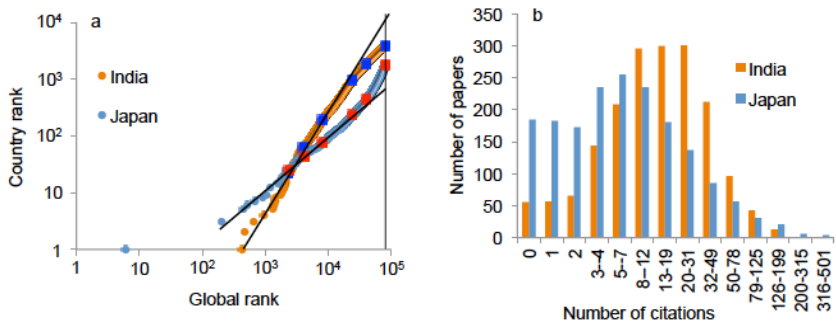

Figure 3. Double rank plots (a) and citation distribution (b) of the publications from India and Japan in graphene. Square symbols in (a) represent the position of papers in top percentiles: 3, 5, 10, 30, 50, and 100, from bottom to top. In (b), the number of papers from India, 3,794, has been scaled down to the number of papers from Japan, 1,788

Despite the described differences, in the two research topics and 16 countries depicted in Tables 2 and 3, the well-known strong relationship between $P_{top\ 10\%}/P$ and the mean number of citations (MNC) (Perianes-Rodriguez & Ruiz-Castillo, 2016; Waltman et al., 2012) holds (Pearson correlation coefficients of 0.97 in graphene and 0.98 in solar cells, $P < 0.001$ in both cases).

*3.2. Basic description of journals*

To further investigate the complexity of citation distributions and their effect on research assessments, I studied journals' publications in a selection of 10 research topics (Table 1), paying special attention to the proportion of uncited papers. As a general rule, although the number of journals publishing papers on a certain topic is very high, the number of journals publishing many papers is much lower. For example, in the field of graphene over the four years of this study, I retrieved 82,757 papers published in 1,648 journals, but only 250 journals published more than 50 papers, 135 journals published more than 100 papers, and 31 journals published more than 500 papers. Notably, 406 journals published only one paper.

In contrast to countries where the absence of uncited papers is infrequent or perhaps never occurs (Tables 2 and 3), many journals in the research topics studied here do not publish papers that are not cited, or they occur at a very low proportion (Table 4). Many of these journals have high JIFs, such as the *Journal of the American Chemical Society* or *ACS Nano* in technological fields, or *Science* or *Cell* in biomedical fields (JIFs in the range of 15–25). However, there are also journals with lower impact factors that either have no uncited papers or have a very low proportion of them. *Scientific Reports* (JIF, 4.0) is an example of these journals. Interestingly, in many of these journals, goodness-of-fit tests support a lognormal citation distribution (Table 4 shows the results of the Kolmogorov-Smirnov test of the log-transformed data). As expected (Perianes-Rodriguez, 2016; Waltman, 2012), the mean number of citations (MNC) in the publication and citation windows of this study (four years in both cases) is highly correlated with the JIF (Pearson coefficient 0.84; $P < 0.001$).

Overall, in technological topics, I found that many journals with a significant proportion of uncited papers have the word "applied" in their names or are published by well-known technological institutions. For example, in the first case, the *Journal of Applied Physics* and *Applied Physics Letters*; and in the second case, several journals published by the Institute of Electrical and Electronics Engineers (IEEE). Journals with a high proportion of uncited papers are less frequent in biomedicine, even in journals dealing with technical applications. For example, in the topic of stem cells, the number of uncited papers in the journal *Cytotherapy* is only 3%.

Table 4. Basic description of journal's publications in four research topics, including goodness-of-fit to lognormal distribution.

| Topic | Journal | JIF[a] | P[b] | P$_0$[c] (%) | MNC[d] | KS[f] *P*-value |
|---|---|---|---|---|---|---|
| Semiconductors | Journal of the American Chemical Society | 14.7 | 437 | 0 | 85.8 | > 0.15 |
| Semiconductors | Advanced Materials | 25.8 | 396 | 0 | 78.7 | > 0.15 |
| Semiconductors | Angewandte Chemie-International Edition | 13.0 | 262 | 0.4 | 77.0 | > 0.15 |
| Semiconductors | ACS Nano | 14.6 | 584 | 0.3 | 59.7 | > 0.15 |
| Semiconductors | ACS Applied Materials & Interfaces | 8.5 | 1120 | 1.1 | 28.6 | 0.01 |
| Semiconductors | Scientific Reports | 4.0 | 1101 | 1.1 | 20.1 | < 0.01 |
| Solar Cells | Journal of the American Chemical Society | 14.7 | 486 | 0 | 121.0 | 0.05 |
| Solar Cells | Advanced Materials | 25.8 | 653 | 0 | 103.9 | < 0.01 |
| Solar Cells | Nano Letters | 12.3 | 460 | 0.2 | 73.5 | > 0.15 |
| Solar Cells | Advanced Energy Materials | 24.9 | 665 | 0 | 52.4 | > 0.15 |
| Solar Cells | Applied Energy | 8.4 | 453 | 0.2 | 40.4 | < 0.01 |
| Solar Cells | Scientific Reports | 4.0 | 805 | 0.6 | 26.9 | < 0.01 |
| Stem cells | Science | 41.1 | 112 | 0 | 190.8 | > 0.15 |
| Stem cells | Nature | 43.1 | 279 | 0 | 180.5 | 0.01 |
| Stem Cells | Cell | 36.2 | 167 | 0 | 168.3 | > 0.15 |
| Stem cells | Nature Communications | 11.9 | 691 | 0 | 51.6 | > 0.15 |
| Stem cells | Blood | 16.6 | 673 | 0.1 | 46.9 | 0.08 |
| Stem cells | Development | 5.8 | 483 | 0.4 | 29.0 | 0.02 |
| Lithium batteries | Advanced Materials | 25.8 | 274 | 0 | 149.1 | > 0.15 |
| Lithium batteries | Journal of the American Chemical Society | 14.7 | 159 | 0 | 139.8 | > 0.15 |
| Lithium batteries | Advanced Energy Materials | 24.9 | 434 | 0 | 93.9 | < 0.01 |
| Lithium batteries | Nano Letters | 12.3 | 313 | 0 | 91.5 | > 0.15 |
| Lithium batteries | Nano Energy | 15.6 | 479 | 0 | 63.5 | < 0.01 |
| Lithium batteries | Scientific Reports | 4.0 | 410 | 0 | 33.7 | 0.14 |

[a] Journal Impact Factor; [b] number of publications; [c] number of uncited papers; [d] Mean Number of Citations; [f] Kolmogorov-Smirnov test of log-transformed number of citations

I also found a group of journals in all topics whose main characteristic is that their most cited papers have a low number of citations and that the proportion of uncited papers is high. These journals may be highly specialized or published in countries that are developing their research systems. In some other cases, they are open-access journals that might belong to the category of predatory journals (Beall, 2012).

Table 5. Proportion of uncited papers in the same journals of publications in the research topics of lithium batteries and solar cells.

|  | Lithium batteries | | Solar cells | |
| --- | --- | --- | --- | --- |
| Journal | P[a] | $P_0$[b](%) | P | $P_0$(%) |
| Advanced Materials | 274 | 0.00 | 653 | 0.00 |
| Journal of the American Chemical Society | 159 | 0.00 | 486 | 0.00 |
| Advanced Energy Materials | 432 | 0.00 | 665 | 0.00 |
| NANO Letters | 313 | 0.00 | 458 | 0.22 |
| Applied Energy | 193 | 0.00 | 453 | 0.22 |
| NANO Energy | 479 | 0.00 | 440 | 0.45 |
| Scientific Reports | 410 | 0.00 | 805 | 0.62 |
| Journal of Materials Chemistry A | 2095 | 0.14 | 1681 | 0.71 |
| ACS Applied Materials & Interfaces | 1416 | 0.14 | 1799 | 1.11 |
| Chemistry of Materials | 529 | 0.19 | 549 | 0.36 |
| Journal of Physical Chemistry C | 624 | 0.48 | 1549 | 2.52 |
| Journal of Power Sources | 2813 | 0.50 | 501 | 1.80 |
| Chemical Communications | 352 | 0.57 | 434 | 0.92 |
| Physical Chemistry Chemical Physics | 514 | 0.58 | 927 | 1.73 |
| Electrochimica Acta | 2382 | 0.67 | 689 | 3.48 |
| Journal of Alloys and Compounds | 839 | 1.07 | 644 | 3.11 |
| RSC Advances | 1634 | 2.14 | 1941 | 3.66 |
| Journal of the Electrochemical Society | 1339 | 2.39 | 206 | 8.74 |
| Journal of Materials Science: Materials in Electronics | 122 | 3.28 | 705 | 11.21 |
| Materials Letters | 411 | 3.41 | 424 | 6.60 |

[a] Number of papers; [b] uncited papers

Although the causes for the different proportions of uncited papers were beyond the scope of this study, a noticeable characteristic in the journals studied is that the proportions of uncited papers in a certain journal in different topics reflect the different proportions of global uncited papers in these topics. Table 5 records the proportion of uncited papers in the same 20 journals in the topics of lithium batteries and solar cells: 3.1% and 7.6% of uncited papers, respectively. The two percentages of uncited papers in the two topics across journals are highly correlated (Pearson coefficient 0.92; $P < 0.001$), and the ratio between these percentages is almost identical to that existing in the global papers in the two topics.

Table 6. Basic description of journal's publications and percentile ratios in the research topic of graphene.

| Journal | JIF[a] | P[b] | $P_0$[c] (%) | MNC[d] | $P_{top10\%}/P$ | $P_{top5\%}/P_{top50\%}$ | $P_{top3\%}/P_{top30\%}$ | $P_{top1\%}/P_{top10\%}$ |
|---|---|---|---|---|---|---|---|---|
| Advanced Materials | 25.8 | 809 | 0.25 | 129.3 | 0.604 | 0.462 | 0.361 | 0.249 |
| Nature Communications | 12.1 | 558 | 0.00 | 129.1 | 0.529 | 0.409 | 0.347 | 0.268 |
| Journal of the American Chemical Society | 14.7 | 285 | 0.00 | 122.2 | 0.519 | 0.386 | 0.330 | 0.277 |
| Advanced Functional Materials | 16.8 | 626 | 0.00 | 91.8 | 0.450 | 0.309 | 0.259 | 0.160 |
| ACS Nano | 14.6 | 1115 | 0.09 | 82.6 | 0.383 | 0.284 | 0.249 | 0.164 |
| Nano Letters | 11.2 | 956 | 0.21 | 64.8 | 0.306 | 0.227 | 0.189 | 0.106 |
| ACS Applied Materials & Interfaces | 8.8 | 2837 | 0.14 | 49.0 | 0.231 | 0.129 | 0.088 | 0.038 |
| ACS Sustainable Chemistry & Engineering | 7.6 | 346 | 0.29 | 47.8 | 0.223 | 0.124 | 0.066 | 0.013 |
| Journal of Materials Chemistry A | 11.3 | 2646 | 0.11 | 43.5 | 0.190 | 0.087 | 0.053 | 0.026 |
| Nanoscale | 6.9 | 1933 | 0.52 | 34.2 | 0.130 | 0.077 | 0.050 | |
| Journal of Power Sources | 8.2 | 1024 | 0.39 | 33.6 | 0.113 | 0.038 | 0.032 | |
| Carbon | 8.8 | 2339 | 1.20 | 30.9 | 0.111 | 0.081 | 0.066 | |
| Scientific Reports | 4.0 | 2008 | 0.90 | 29.4 | 0.092 | 0.056 | 0.048 | |
| 2D Materials | 5.5 | 414 | 0.72 | 27.8 | 0.075 | 0.048 | | |
| Electrochimica Acta | 6.2 | 2217 | 0.13 | 24.9 | 0.042 | 0.016 | 0.012 | |
| Journal of Physical Chemistry C | 4.2 | 1310 | 1.83 | 22.2 | 0.059 | 0.039 | 0.030 | |
| Physical review B | 3.6 | 2027 | 3.90 | 15.3 | 0.033 | 0.049 | 0.047 | |
| RSC Advances | 3.1 | 5127 | 2.05 | 15.7 | 0.017 | 0.010 | | |
| Applied Physics Letters | 3.6 | 1188 | 3.70 | 14.1 | 0.016 | 0.017 | | |
| Journal of Applied Physics | 2.3 | 682 | 6.01 | 9.4 | 0.007 | | | |

[a] Journal Impact Factor; [b] number of papers; [c] uncited papers; [d] Mean Number of Citations; $P_{top\,x\%}$, number of papers in top percentile x

## 3.3. Deviations from the rank power law: journals

In all cases studied, deviations of the global ranks of journals' papers from a power law are frequent and show many similarities. Table 6 depicts the basic analysis of 20 journals with JIFs ranging from 26 to 2 in the research topic of graphene. This table shows the top percentile ratios for testing the ideal rank power law ($P_{top\ 10\%}/P = P_{top\ 5\%}/P_{top\ 50\%} = P_{top\ 3\%}/P_{top\ 30\%} = P_{top\ 1\%}/P_{top\ 10\%}$). This method can only be applied to journals where the numbers of papers in the top 5% or 3% of most cited papers are statistically significant, which occurs only in journals with high JIFs or many publications. The first conclusion drawn from the data is that, as a general rule, journals with JIFs above approximately 4.0 notably deviate from the rank power law. Most of these journals do not have uncited papers, or their proportion is very low. Generally, the proportion of uncited papers increases as the JIF decreases. Notably, the well-known relationship between $P_{top\ 10\%}/P$ and MNC (Perianes-Rodriguez & Ruiz-Castillo, 2016; Waltman et al., 2012) applies to journals (Pearson correlation coefficient, 0.99, $P < 0.001$).

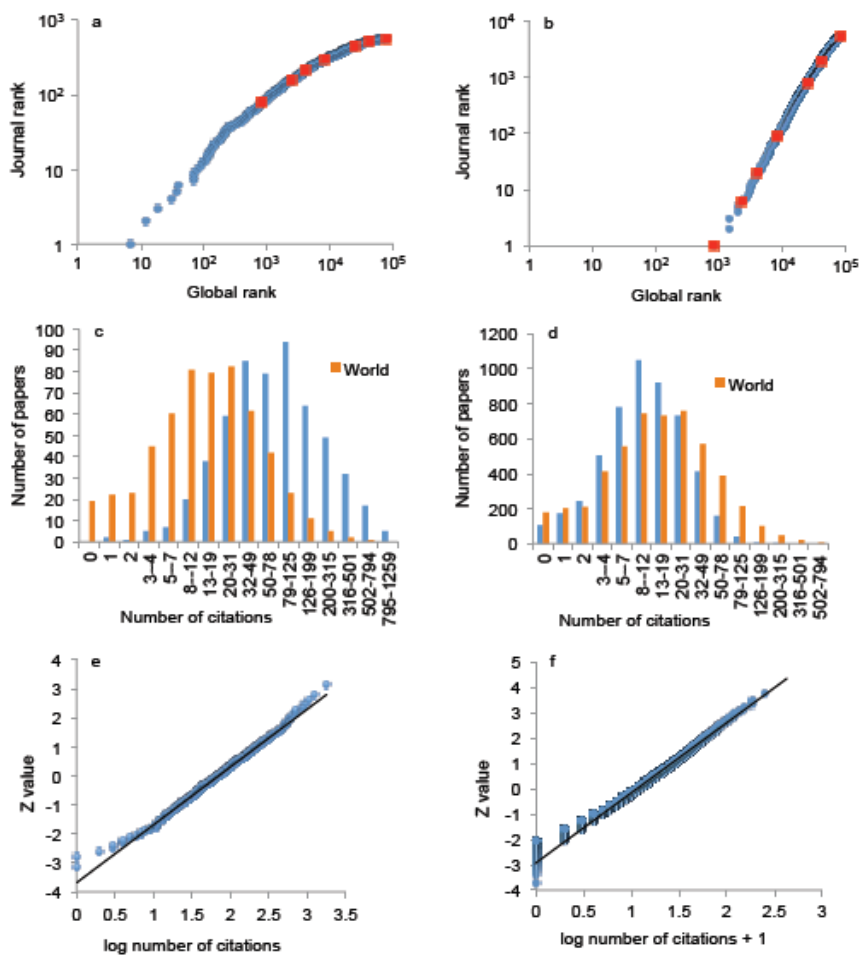

Figure 4. Publications on graphene in journals *Nature Communications* (a, c, e) and *RSC Advances* (b, d, f) and. Double rank plots (a, b), citation distributions (c, d), and normal probability plots (e, f). In (a) and (b), square symbols represent the position of papers in top percentiles: 1, 3, 5, 10, 30, 50, and 100, from bottom to top. Histograms in (c) and (d) include world publications, which have been scaled down to the number of publications in the journals: from 82,757 to 5,127, and 558, respectively

Figure 4 provides more detailed information from two journals, *Nature Communications* and *RSC Advances*, with high and medium JIFs of 12.1 and 3.1, respectively. This figure includes the double rank plot, the citation distribution using logarithmic binning in comparison with the global distribution (downscaled to the number of papers in the journal), and the normal probability plot of the log-transformed number of citations. In reference to the global citation distribution, the distribution in *Nature Communications* is notably shifted to the right, without uncited publications and with a much lower proportion of papers with 1 or 2 citations. Probably as a consequence, the double rank plot is not linear, except for the top 1% cited papers. The normal probability plot of the log-transformed number of citations shows only slight deviations from a straight line, and the Kolmogorov-Smirnov goodness-of-fit test ($P > 0.15$) indicates that citations can be modeled according to a lognormal distribution. In contrast, the citation distribution in *RSC Advances* is similar to the global one in the papers with 0, 1, and 2 citations but is slightly shifted to the left in the rest of the distribution. The double rank plot is a straight line with an insignificant deviation in the 50% least cited papers. The normal probability plot of the log-transformed number of citations plus 1 indicates that the lowly cited papers notably deviate from a lognormal citation distribution.

The double rank plot of *Nature Communications* (Figure 4) represents the extreme of the deviations of journals' plots from a straight line, deviations that disappear in journals with JIFs around 4 or lower. This implies that in many journals, evaluations based on the number of papers and a single indicator do not describe the data. For example, Figure 5 shows the comparison of *Advanced Materials* with *Nanoscale* (JIFs, 25.8 and 6.9, respectively) and of *Electrochimica Acta* with *RSC Advances* (JIFs, 6.2 and 3.1, respectively). The first case (panels a and b) shows that *Advanced Materials* notably exceeds *Nanoscale* in the probability of publishing very highly cited papers, but the obvious difference cannot be accurately described with a single parameter. The number of papers is higher in *Nanoscale* (1,933 versus 809), while $P_{top\ 10\%}$ and $P_{top\ 1\%}$ are higher in *Advanced Materials* (489 versus 251, and 122 versus 5). Furthermore, if we calculate the ratios between these parameters ($P_{top\ 10\%}/P$, $P_{top\ 1\%}/P$, and $P_{top\ 1\%}/P_{top\ 10\%}$), the differences between the two journals are so disproportioned (0.6 versus 0.12, 0.15 versus 0.003, 0.25 versus 0.02) that it would be illogical to select only one of them to describe the journals' difference.

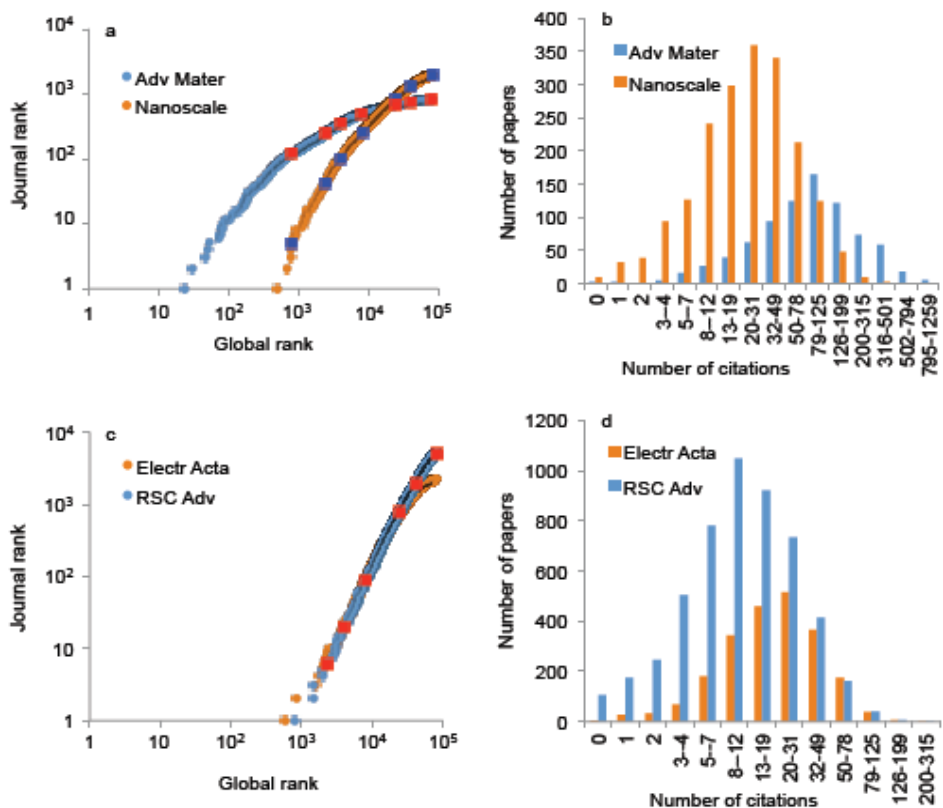

Figure 5. Publications on graphene in journals *Advanced Matter, Nanoscale, Electrochimica Acta,* and *RSC Advances*. Double rank plots (a, c) and citation distributions (b, d). In (a) and (c), square symbols represent the position of papers in top percentiles: 1, 3, 5, 10, 30, 50, and 100, from bottom to top (in panel c, $P_{top\ 1\%}$ is not marked).

In the second case (panels c and d), in the number of publications, *RSC Advances* more than doubles *Electrochimica Acta* (5127 versus 2217, respectively), but the double rank plots of the top 30% cited papers are almost identical in both journals. Consequently, $P_{top\ 10\%}$ has almost the same value in the two journals (785 and 743, respectively). If we define the efficiency in the rank range of the top 30% most cited papers, the two journals are equal, but the $P_{top\ 10\%}/P$ ratio of *Electrochimica Acta* doubles that of *RSC Advances* (0.34 versus 0.15). The important difference in "efficiency" between the two journals is in the lowly cited papers. In the highly cited papers, the two journals are identical. Again, it is doubtful that these differences can be described with a single indicator.

These results with graphene suggest that differences in the proportion of uncited papers between journal and global publications affect the double rank plots. But graphene is a research topic with a low proportion of uncited papers, which raises the question of whether in a topic with a higher proportion of uncited papers (e.g. semiconductors) the conclusions would be different. To investigate this issue, I compared the double rank plots of the same two journals, *ACS Applied Materials and Interfaces* and *Scientific Reports* (JIFs, 8.8 and 4.0; uncited papers, 0.14% and 0.90%; Table 6), in the two research topics of graphene and semiconductors, which have different proportions of uncited papers (3.8% and 10.8%, respectively; Table 1). The results (Figure 6) show that the plot of *Scientific Reports* in semiconductors deviates more from linearity (increased curvature) than in graphene, but that this does not occur in *ACS Applied Materials and Interfaces*. These results suggest that the proportion of uncited papers in the topic contributes to deviations of the double rank plot from linearity but that this contribution is negligible in journals with high JIF, which show high deviations even in topics with a low proportion of uncited papers.

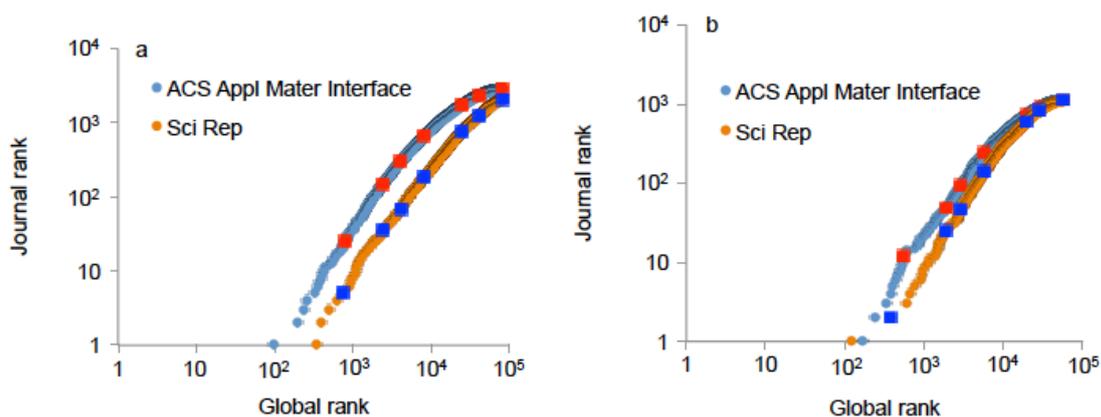

Figure 6. Double rank plots of publications in graphene (a) and semiconductors (b) in the journals *ACS Applied Materials and Interfaces* and *Scientific Reports*. Square symbols represent the position of papers in top percentiles: 1, 3, 5, 10, 30, 50, and 100, from bottom to top

*3.5. Log-log double rank plots can have positive and negative curvatures*

The results depicted in Figure 3 reveal that the log-log double rank plot corresponding to the lower tail can have downward or upward concavity. Downward concavity can be easily studied in journals, but not upward concavity. The former appears when the journal has a lower proportion of uncited papers than the global publications, and the latter should appear in journals with a higher proportion of uncited papers than the global publications. These journals exist, but in these journals, most of their publications are lowly cited. Consequently, all the papers are grouped in the lower tail of the global citation distribution, and in these cases, the double rank plot does not deviate or deviates very little from a power law.

Table 7. Percentile ratios of publications from countries on the research topics of graphene and solar cells.

| Country | Graphene | | Solar cells | |
|---|---|---|---|---|
| | $P_{top\ 50\%}/P_{top\ 100\%}$ | $P_{top\ 5\%}/P_{top\ 10\%}$ | $P_{top\ 50\%}/P_{top\ 100\%}$ | $P_{top\ 5\%}/P_{top\ 10\%}$ |
| Australia | 0.59 | 0.62 | 0.56 | 0.44 |
| Canada | 0.49 | 0.44 | 0.57 | 0.45 |
| China | 0.51 | 0.50 | 0.48 | 0.47 |
| EU | 0.36 | 0.49 | 0.51 | 0.41 |
| Germany | 0.38 | 0.57 | 0.55 | 0.49 |
| India | 0.48 | 0.33 | 0.46 | 0.36 |
| Japan | 0.24 | 0.55 | 0.33 | 0.45 |
| Singapore | 0.63 | 0.60 | 0.58 | 0.56 |
| South Korea | 0.46 | 0.41 | 0.38 | 0.44 |
| UK | 0.50 | 0.47 | 0.58 | 0.57 |
| USA | 0.47 | 0.62 | 0.58 | 0.56 |

$P_{top\ x\%}$, number of papers in top percentile x

Therefore, I further investigated the issue in countries. Table 7 depicts the $P_{top\ 50\%}/P$ and $P_{top\ 5\%}/P_{top\ 10\%}$ ratios in 11 countries (for statistical robustness the number of countries is reduced with reference to Tables 2 and 3; in all cases, $P_{top\ 5\%}$ > 22) in the topics of graphene and solar cells. The $P_{top\ 50\%}/P$ ratio corresponds to the maximum curvature in the log-log double rank plots (Figures 5 and 6), and the $P_{top\ 5\%}/P_{top\ 10\%}$ ratio corresponds to the top 10% tail, which in countries fit the power law (Figure 3). In countries behaving as ideal systems, the $P_{top\ 50\%}/P$ and $P_{top\ 5\%}/P_{top\ 10\%}$ ratios are equal.

Overall, in Table 7, in 11 cases out of 22, the ratios are equal (deviations < 15%), while in the other cases, $P_{top\ 50\%}/P$ is larger or lower than $P_{top\ 5\%}/P_{top\ 10\%}$, reproducing India and Japan in Figure 3. In graphene, in three countries: Germany, Japan, and the USA, $P_{top\ 50\%}/P$ is lower than $P_{top\ 5\%}/P_{top\ 10\%}$, and in one country, India, $P_{top\ 50\%}/P$ is larger than $P_{top\ 5\%}/P_{top\ 10\%}$. In solar cells, in two countries: Japan and South Korea, $P_{top\ 50\%}/P$ is lower than $P_{top\ 5\%}/P_{top\ 10\%}$, and in three countries: Australia, Canada, and India, $P_{top\ 50\%}/P$ is larger than $P_{top\ 5\%}/P_{top\ 10\%}$. In the EU, $P_{top\ 50\%}/P$ is lower than $P_{top\ 5\%}/P_{top\ 10\%}$ in graphene and larger in solar cells.

## 4. DISCUSSION

The aim of this study was to explore when statistical data that can be easily obtained, such as $P_{top\ 10\%}/P$ or $P_{top\ 1\%}/P$, can or cannot be used for research assessment. It is firmly established that when the global ranks of local publications follow an ideal power law, the assessment with these statistical data is accurate (Rodríguez-Navarro & Brito, 2021); the challenge is that deviations occur (Rodríguez-Navarro, 2024b). To address this challenge, the use of journals provides many advantages. Probably no country or institution can have a research success similar to that of the papers published in *Nature* or *Science*, but these and similar journals provide clues for research evaluation that would be impossible to obtain with countries and institutions. Another hallmark of this study is the use of research topics. Different topics that are published in the same journals or are together in the same research field may have very different citation distributions, especially in the proportions of uncited papers (Table 1), which is a crucial datum in the analysis of research indicators. The effects of these uncited papers on indicators are more difficult to study in the mix of many topics, where internal compensations occur, than when the topics are studied independently.

*4.1. Conformity with the ideal double rank power law*

Analyses of percentile ratios across countries and research topics demonstrate that conformity with the ideal power law ($P_{top\ 10\%}/P = P_{top\ 5\%}/P_{top\ 50\%} = P_{top\ 3\%}/P_{top\ 30\%} = P_{top\ 1\%}/P_{top\ 10\%}$) is frequent, but deviations from the power law, either with increasing or decreasing patterns of these serial ratios, are also frequent (Tables 2 and 3). For example, the EU shows increasing ratios in graphene but decreasing in solar cells; the USA shows increasing ratios in graphene but constant in solar cells; India shows increasing ratios in both topics; and Japan shows decreasing ratios in both topics. In these conditions, the comparison of countries with a single per-publication indicator of efficiency (e.g., $P_{top\ 10\%}/P$) is uncertain. According to double rank plots, countries' comparisons will be reliable when deviations from the power law are not important (Figure 2) but impossible to perform with a single indicator when the deviations are important (Figure 3).

Deviations from the power law appear when there are significant differences between countries' and global citation distributions (Rodríguez-Navarro, 2024b), especially with reference to the proportion of uncited papers (Tables 1–3). Possibly, in global and countries' distributions, there is always an excess of uncited papers with reference to a lognormal distribution (Rodríguez-Navarro, 2024b). Figure 1 depicts the normal probability plot of the log-transformed number of citations (plus 1) in two topics, which suggests that most of the citation distribution could be modeled according to a lognormal distribution but that the lower tail has an excess of lowly cited papers. However, it is worth noting that discretization of a continuous lognormal distribution of random numbers with a low $\mu$ parameter also produces an apparent excess of zeros because, in such continuous series, many numbers are less than 0.5. This implies that for testing the lognormal distribution of citations, normal probability plots of the log-transformed number of citations must be performed in parallel with other types of analysis.

In graphene, the proportion of uncited papers in *Nature Communications*, *Advanced Materials*, *Nanoscale*, and *Electrochimica Acta* is lower than in global papers, and the right extreme of the log-log double rank plot shows downward concavity, as shown in Figures 4, 5, and 6. The analysis of these figures suggests that the deviation of the double rank plots, namely, the degree of the curvature, depends on the difference in the lowly cited papers (zero, one, and two citations) between the journal and global citation distributions. However, the comparison of *ACS Applied Materials & Interfaces* and *Scientific Reports* in the topics of graphene and semiconductors (Figure 6) suggests that maximum deviation seems to have a limit. In *ACS Applied Materials & Interfaces*, the curvature is similar in graphene and semiconductors, although there is a large difference in the proportion of uncited papers in these two topics (Table 1), while the proportion of uncited papers in both topics in this journal is similar (0.14% and 0.11%, Tables 6 and 4, respectively).

Widely used percentile indicators (e.g., $P_{top\ 10\%}/P$) are not absolute measures of research success but relative measures because they are obtained by comparing the research of countries and institutions with global research. The results of this study indicate that when the citation distribution of a country or institution is very different from the global distribution, especially in the proportion of uncited papers, the double rank plot deviates from a power law. In these cases, research assessments cannot be performed with easy-to-obtain statistical data.

*4.2. Not less than three indicators are necessary for the research assessment of countries*

The translation of the conclusions reached with journals to countries warns against using the number of publications and another single indicator, e.g., $P_{top\ 10\%}/P$, for the evaluation of countries and institutions. Most current evaluations use this method, which may be highly misleading in some cases. A reasonable conjecture is that the method undervalues countries with high-technology industries (Japan, Germany, South Korea, USA, etc.) in those technological topics where they have global influence, and overvalues countries where there is academic research that is not being applied by a competitive industry. This latter case is probably more frequent in research-leading universities (Rodríguez-Navarro, 2024b). Most of these deviations may be predicted from the proportion of uncited papers. In countries and institutions with research that is oriented toward technological improvements, the proportion of uncited papers is higher than the mean shown by global publications. Conversely, in countries and institutions dominated by academic research, the proportion of uncited papers is lower than the average shown by global publications. Consequently, in both cases, using P (total number of publications) to calculate the $P_{top\ 10\%}/P$ or $P_{top\ 1\%}/P$ ratios and subsequently use the results in research assessments will be misleading in many cases.

These considerations about academic and applied research, and the proportion of lowly cited papers, raise an interesting question about the evaluations of research efficiency. In principle, the ratio between applied and academic research is a structural characteristic of research systems that is not necessarily linked to the efficiency of each type of research. This point of view leads again to the simple conclusion that research efficiency cannot be measured with a single size-dependent indicator, which is an unquestionable datum, divided by the total number of papers. More research is necessary to find the most appropriate indicators. However, the indicators used in Table 7 can be obtained easily and might be used to characterize the bulk of research. A different issue is the contribution to pushing the boundaries of knowledge, which might depend on research elites that may not be visible when the bulk of papers is considered. This contribution can be easily measured from the ranks of the most cited papers (Rodríguez-Navarro & Brito, 2024). However, this is a size-dependent parameter that should be normalized to a size-independent measure to compare countries or institutions, which is a challenge if P cannot be used for this purpose. Abramo and D'Angelo (2016a, 2016b) criticized the use of common size-independent indicators from the point of view of research performance. The issue raised in the present study is different and refers to the misleading use of per-publication indicators as a measure of efficiency.

These results imply that a minimum of four indicators, either parametric or nonparametric, are necessary to describe a research system: (i) the size; (ii) the lower tail (the 50% least cited papers), which informs about the comparison of the research system in terms of lowly cited papers; (iii) the extreme of the upper tail, in order to find out the contribution to pushing the boundaries of knowledge (Rodríguez-Navarro & Brito, 2024); and (iv) an indicator for the 50% most cited papers. This last indicator might be used to determine the efficiency of the portion of the system that has an academic orientation, pursuing pushing the boundaries of knowledge.

## 5. Conclusions and implications

The use of a single size-independent indicator to describe the research output of countries and institutions is highly widespread among international and national agencies. The most frequent are $P_{top\ 10\%}/P$ and $P_{top\ 1\%}/P$. This study demonstrates that these assessments and the corresponding rankings are correct in some cases but misleading in others. Consequently, the use of these indicators for research assessments creates drawbacks and confusion that conceal the benefits. The statement of Garfield and Welljams-Dorof (Garfield & Welljams-Dorof, 1992, p. 321): "Government policy-makers, corporate research managers, and university administrators need valid and reliable S&T indicators for a variety of purposes: for example, to measure the effectiveness of research expenditures, identify areas of strength and excellence, set priorities for strategic planning, monitor performance relative to peers and competitors, and target emerging specialties and new technologies for accelerated development" defines a need that currently, more than 30 years after it was written, has not been met. Therefore, every effort should be made to find a solution to this challenge; the results of this study open an avenue of research to reach that solution.